\documentclass[autodetect-engine,dvipdfmx,10pt,a4paper,twocolumn]{article}

\usepackage[linesnumbered,ruled]{algorithm2e}
\usepackage{bm}
\usepackage{amsmath}
\usepackage{amssymb}
\usepackage{amsthm}
\usepackage{color}
\usepackage{dsfont}
\usepackage{booktabs}
\usepackage{framed}
\usepackage{tabularx}
\usepackage{graphicx}

\newenvironment{tsaligned}{%
\begin{equation}\begin{aligned}%
}{%
\end{aligned}\end{equation}%
}
\newenvironment{tsaligned*}{%
\begin{equation*}\begin{aligned}%
}{%
\end{aligned}\end{equation*}%
}

\newcommand{\tsja}[1]{}
\newcommand{\tsnote}[1]{}

\theoremstyle{definition}
\newtheorem{theorem}{Theorem}
\theoremstyle{definition}
\newtheorem{lemma}[theorem]{Lemma}
\theoremstyle{definition}

\theoremstyle{definition}

\theoremstyle{definition}

\theoremstyle{definition}
\newtheorem{definition}[theorem]{Definition}
\theoremstyle{definition}
\newtheorem{proposition}[theorem]{Proposition}
\theoremstyle{definition}
\newtheorem{property}[theorem]{Property}
\theoremstyle{definition}
\newtheorem{example}[theorem]{Example}
\theoremstyle{definition}
\newtheorem{conjecture}[theorem]{Conjecture}
\theoremstyle{definition}

\theoremstyle{definition}

\theoremstyle{definition}
\newtheorem{corollary}[theorem]{Corollary}
\theoremstyle{definition}
\newtheorem{exercise}[theorem]{Exercise}
\theoremstyle{definition}
\newtheorem{simulation}[theorem]{Simulation}

{\endMakeFramed}

\newenvironment{greenleftbar}{%
\MakeFramed {\advance\hsize-\width \FrameRestore}}%
{\endMakeFramed}

\newenvironment{lightgrayleftbar}{%
\MakeFramed {\advance\hsize-\width \FrameRestore}}%
{\endMakeFramed}

\newenvironment{example-waku}
{\begin{lightgrayleftbar}\begin{example}}
{\end{example}\end{lightgrayleftbar}}
\newenvironment{exercise-waku}
{\begin{lightgrayleftbar}\begin{exercise}}
{\end{exercise}\end{lightgrayleftbar}}
\newenvironment{simulation-waku}
{\begin{greenleftbar}\begin{simulation}}
{\end{simulation}\end{greenleftbar}}
\newenvironment{proposition-waku}
{\begin{oframed}\begin{proposition}}
{\end{proposition}\end{oframed}}
\newenvironment{definition-waku}
{\begin{oframed}\begin{definition}}
{\end{definition}\end{oframed}}
\newenvironment{lemma-waku}
{\begin{oframed}\begin{lemma}}
{\end{lemma}\end{oframed}}
\newenvironment{theorem-waku}
{\begin{oframed}\begin{theorem}}
{\end{theorem}\end{oframed}}
\newenvironment{property-waku}
{\begin{oframed}\begin{property}}
{\end{property}\end{oframed}}
\newenvironment{corollary-waku}
{\begin{oframed}\begin{corollary}}
{\end{corollary}\end{oframed}}
\newenvironment{conjecture-waku}
{\begin{oframed}\begin{conjecture}}
{\end{conjecture}\end{oframed}}

\newcommand{\rh}{{\textnormal{h}}}

\newcommand{\rv}{{\textnormal{v}}}

\newcommand{\0}{{\bm{0}}}

\newcommand{\vb}{{\bm{b}}}

\newcommand{\vw}{{\bm{w}}}
\def\x{\bm{x}}

\newcommand{\y}{{\bm{y}}}

\newcommand{\vA}{{\bm{A}}}

\newcommand{\bE}{{\mathbb{E}}}

\newcommand{\vI}{{\bm{I}}}
\newcommand{\cI}{{\mathcal{I}}}

\newcommand{\cN}{{\mathcal{N}}}

\newcommand{\vO}{{\bm{O}}}

\newcommand{\bR}{{\mathbb{R}}}

\newcommand{\X}{{\bm{X}}}

\newcommand{\vlam}{{\bm{\lambda}}}

\DeclareMathOperator{\diag}{diag}

\newcommand{\argmax}{\mathop{\textrm{argmax}}\limits}

\newenvironment{summary}{\textbf{Abstract: }}{}
\newenvironment{keywords}{\textbf{Keywords: }}{}

\newcommand{\tslong}[1]{{#1}}
\newcommand{\tsshort}[1]{{}}

\begin{document}
\title{Asymmetric Tobit analysis for correlation estimation from censored data}

\author{HongYuan Cao${}^{1}$ and Tsuyoshi Kato${}^{1,2}$
\\
\normalsize
${}^1$~Faculty of Science and Technology, Gunma University, \\
\normalsize
Tenjin-cho 1-5-1, Kiryu, Gunma 376-8515, Japan. %
\\
\normalsize
${}^2$~Integrated Institute for Regulatory Science, Waseda University, \\
\normalsize
513 Wasedatsurumakicho, Shinjuku, Tokyo, 162-0041, Japan. %
}
\maketitle

\begin{summary}
Contamination of water resources with pathogenic microorganisms excreted in human feces is a worldwide public health concern. 
Surveillance of fecal contamination is commonly performed by routine monitoring for a single type or a few types of microorganism(s). 
To design a feasible routine for periodic monitoring and to control risks of exposure to pathogens, reliable statistical algorithms for inferring correlations between concentrations of microorganisms in water need to be established. Moreover, because pathogens are often present in low concentrations, some contaminations are likely to be under a detection limit. This yields a pairwise left-censored dataset and complicates computation of correlation coefficients. 
Errors of correlation estimation can be smaller if undetected values are imputed better. To obtain better imputations, we utilize side information and develop a new technique, the \emph{asymmetric Tobit model} which is an extension of the Tobit model so that domain knowledge can be exploited effectively when fitting the model to a censored dataset. 
The empirical results demonstrate that imputation with domain knowledge is effective for this task.

\end{summary}

﻿\section{Introduction}
\label{s:intro}
\begin{figure}[t!]
  \centering
  \footnotesize
  \begin{tabular}{l}
    \includegraphics[width=0.4\textwidth]{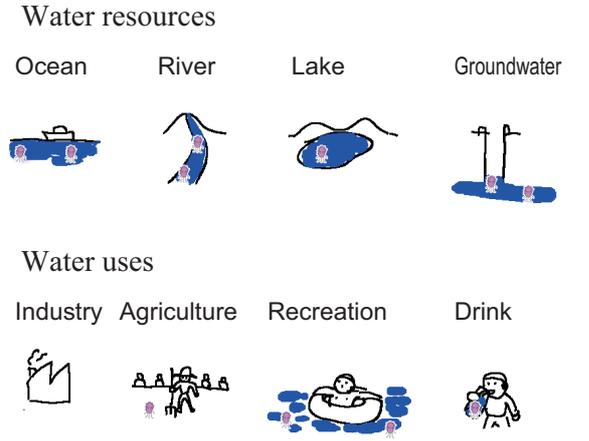}
  \end{tabular}
  \caption{%
  Water resources and uses. Fecal contamination in water resources leads to microbial risk of exposure to waterborne pathogens through various water uses including drinking, recreation, agriculture, and industry. 
  \label{fig:watercontami}
  }
\end{figure}

Contamination of water resources with pathogenic microorganisms excreted in human feces is a public health concern worldwide. 
Contamination of water with several types of pathogenic microorganisms, such as bacteria and viruses, causes diseases in humans. 
Well-known harmful enteric bacteria include \textit{Salmonella}, \textit{Shigella}, and \textit{Escherichia coli} (\textit{E. coli}) O157:H7, while 
 \textit{enterovirus}, \textit{norovirus}, and \textit{rotavirus} are common pathogenic viruses.
Oral ingestion is the primary transmission route of enteric illnesses 
(See Figure~\ref{fig:watercontami}). 
Numerous enteric pathogens remaining in treated wastewater contaminate the environment when they are returned to seawater, rivers, lakes and groundwater~\cite{Dobrowoski08a,Pedrero-AWM10a}. 
Pathogens in seawater condense in shellfish, leading to enteric illnesses transmitted by consumption of raw or undercooked shellfish grown in sewage-polluted seawater~\cite{GenVinLip09}. 
The microbial quality of groundwater tends to be relatively stable due to filtration through layers of soil, although it was reported that in the United States, approximately half of waterborne disease
outbreaks are associated with polluted groundwater~\cite{KraHerCra-MCSS96a}. 
Outbreaks associated with untreated recreational waters in rivers, lakes, and ocean often occur owing to fecal contamination. 
Adequate assessment of microbial water quality is required in order to control public health risks related to exposure to pathogenic microorganisms. 

It is almost impossible to include all pathogens in periodic routine monitoring by checking the contamination level of each pathogenic microorganism. Current measurement technologies consume considerable expense and labor for many pathogens, making routine monitoring of such pathogens prohibitive.  A more feasible approach to controlling public health risk from waterborne pathogens is to routinely test for only a few selected types of pathogens. The common targets of routine monitoring of water quality are harmless indicator microorganisms and physicochemical water qualities. 
Commonly used indicators are \textit{total coliforms}, \textit{fecal coliforms}, \textit{enterococci}, and \textit{F-specific bacteriophage}
~\cite{Boehm2014a,McMinn-lam2017,KorMcMHar18,SedVarMeo18,NapHonIch19,GohSaeGu19}. 
Physicochemical water quality measurements include pH (potential of hydrogen), BOD (biochemical oxygen demand), COD(chemical oxygen demand), SS (suspended solids), DO (dissolved oxygen), TN (total nitrogen), and TP (total phosphorus)~\cite{KatKobOis19}. %
However, concentrations of these indicators and physicochemical data may not necessarily be correlated strongly with the presence of pathogenic microorganisms and may not suffice to assess waterborne infectious risk. Meanwhile, with continuous efforts made by many researchers in the water engineering field, new detection technologies for pathogenic microorganisms in water are being developed~\cite{Noble2005,IWP04}. Establishment of statistical techniques for analyzing pathogenic measurement data~\cite{KatKobIto15a,KatMiuOkaSan13a,ItoKatHasKatIshOkaSan17a,ItoKatTak15a} is expected to enable future advancements in  the design of routine monitoring approaches for pathogen detection. 

\emph{Pearson correlation coefficient} (PCC) is a standard measure in the water engineering field for evaluating the relationship between concentrations of two microorganisms~\cite{XiaotongWen-Sus20}. 
The microorganism concentrations that have higher correlation coefficient with concentrations of another target microorganism are more effective in predicting concentrations of the target microorganism. 
Computation of the PCC for indicator--pathogen pairs tends to be a challenge in attempting to measure the relationship between concentrations of two pathogenic microorganisms in water.
The difficulty is caused due to the existence of detection limits, which are not included in the standard setting of statistical analysis. 
Many pathogens are present in low concentrations. To detect the few individuals of such a pathogen, a large volume of water must be sampled, which burdens procedures for periodic routine monitoring with a heavy workload. For monitoring
based on realistic volume sampling, samples of pathogen concentrations are usually left-censored data~\cite{KatKobIto15a,KatMiuOkaSan13a}. 
A na\"{i}ve approach to estimation of PCC between such data is to discard undetected data and to compute PCC only from data pairs in which both pathogens were detected. However, this approach suffers from a severe disadvantage in that commonly detected data amounts to be too low to infer correlation. 
Thus, reliable algorithms for inferring PCC from censored data need to be established to ensure safe and sustainable water resources for human societies on Earth. 

In this study, we investigated the performance of several methods for inferring PCC between censored concentration data of two microorganisms in water. We examined a more sophisticated approach than the aforementioned na\"{i}ve method, exploiting side information to impute undetected concentrations before computing PCC. We fitted a Tobit model~\cite{Amemiya1984} to the censored data and imputed the undetected data with expected values based on  the model. Then, more complete data can be used to infer PCC. The estimation accuracy of this approach depends on the imputation accuracy. To improve the imputation accuracy, we consider exploitation of domain knowledge. For water quality data, the signs of the correlations between any pair of two variates are known in advance. A third approach utilizes this knowledge by introducing the \emph{asymmetric normal distribution}~\cite{KatOmaAso02a} as the prior for the regression coefficients of the Tobit model. 

Another technical contribution of this study is the discovery of an efficient algorithm for fitting the Tobit model with the asymmetric normal prior. An expectation-maximization (EM) algorithm can be used for fitting the classical Tobit model. Each iteration of the EM algorithm consists of an E-step and an M-step. If the prior of the regression coefficients is the ordinary normal distribution, an M-step can be performed by simply solving a linear system. In general, M-steps tend to be challenging if the prior is changed. In this study, we found that M-steps can still be performed efficiently even if the asymmetric normal distribution is adopted as the prior of the regression coefficients. 

This paper is organized as follows. The next section provides a review of three fundamental tools as preliminaries to the later sections: the PCC, a Tobit model, and the nonnegative least square. In Section~\ref{s:pccest}, we introduce three approaches for correlation analysis: a na\"{i}ve approach, a classical Tobit approach, and an asymmetric Tobit approach. In Section~\ref{s:asym}, we present a new algorithm for fitting the asymmetric Tobit model to censored data. In Section~\ref{s:exp}, simulation results are reported. The final section summarizes and concludes the contributions of this study.

\begin{figure}[t!]
  \centering
  \begin{tabular}{lll}
    (a) Na\"{i}ve
    &
    (b) Classical Tobit
    &
    (c) Asymmetric Tobit
    \\
    \includegraphics[height=2cm]{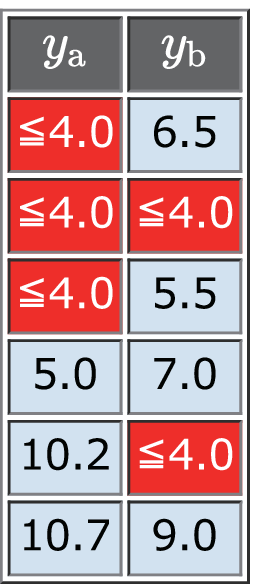}            
    &
    \includegraphics[height=2cm]{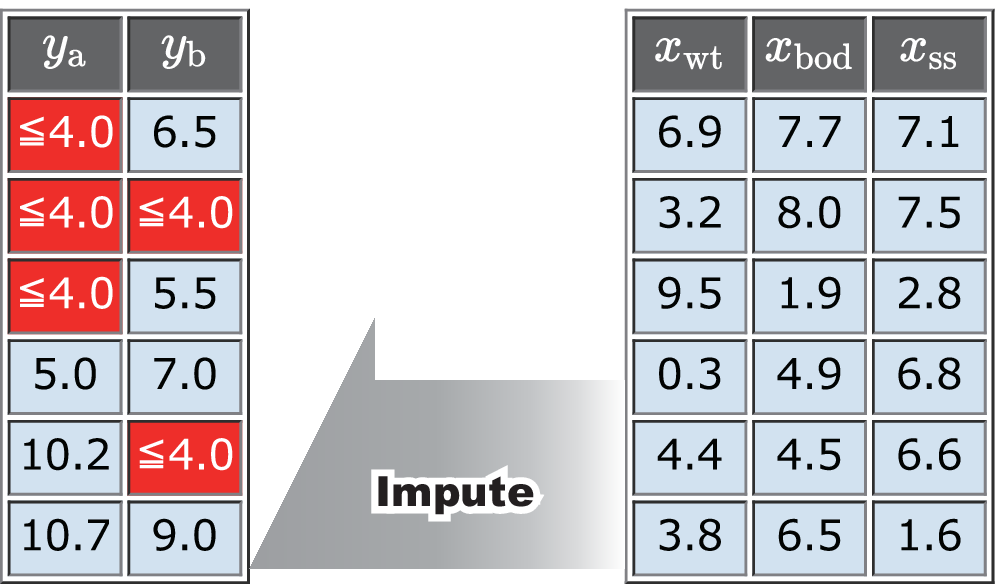}      
    &
    \includegraphics[height=2cm]{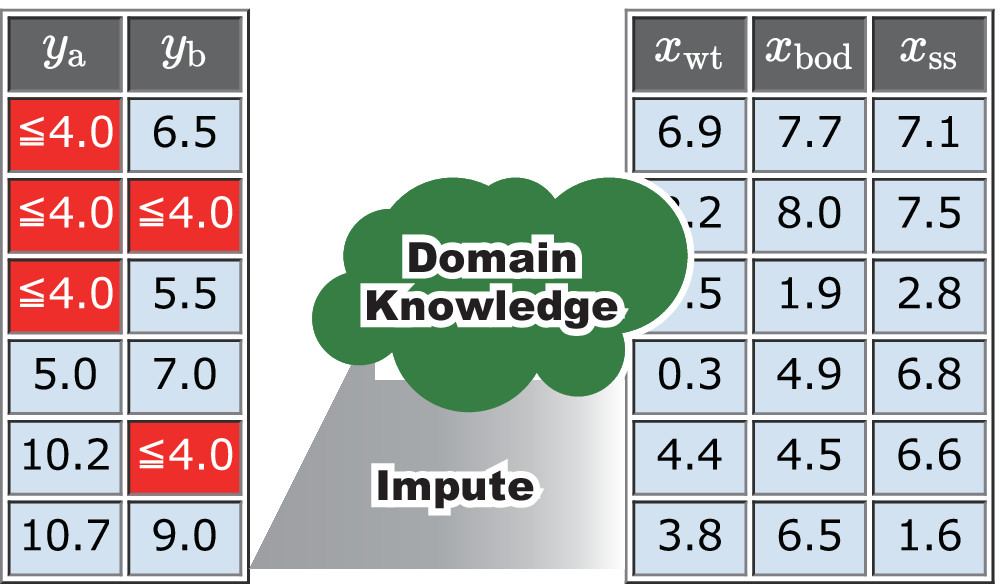}
    \\ 
  \end{tabular}
  \caption{%
  Three approaches for correlation analysis. The targets to be analyzed are censored. (a) Na\"{i}ve approach computes the correlation only from commonly available entries. (b) Classical Tobit approach imputes the missing entries using side information before correlation computation. (c) Asymmetric Tobit approach exploits domain knowledge to improve the imputations. %
  \label{fig:3meth}
  }
\end{figure}

\section{Preliminaries}\label{s:pre}
\subsection{Pearson correlation coefficient}
\label{ss:pcc}
PCC is a statistic for paired data: $(y_{1,\text{a}}, y_{1,\text{b}}), \dots, (y_{n,\text{a}}, y_{n,\text{b}})\in\bR\times\bR$. 
The definition of PCC is given by 
\begin{tsaligned}
  R(\y_{\text{a}},\y_{\text{b}})
  :=
  \frac{\sum_{i=1}^{n}(y_{i,\text{a}}-\bar{y}_{\text{a}})(y_{i,\text{b}}-\bar{y}_{\text{b}})}
  {\sqrt{\sum_{i=1}^{n}(y_{i,\text{a}}-\bar{y}_{\text{a}})^{2}}
  \sqrt{\sum_{i=1}^{n}(y_{i,\text{b}}-\bar{y}_{\text{b}})^{2}}}
\end{tsaligned}      
where 
$\y_{n,\text{a}}:=\left[y_{1,\text{a}},\dots,y_{n,\text{a}}\right]^\top$, 
$\y_{n,\text{b}}:=\left[y_{1,\text{b}},\dots,y_{n,\text{b}}\right]^\top$, 
\begin{tsaligned}
  & \bar{y}_{\text{a}} 
  := 
    \frac{1}{n}  
  \sum_{i=1}^{n}
  y_{i,\text{a}}
  \;\;\text{and}\;
  \bar{y}_{\text{b}} 
  := 
  \frac{1}{n}
  \sum_{i=1}^{n}
  y_{i,\text{b}}. 
\end{tsaligned}

\subsection{Tobit analysis}
\label{ss:symtbt}
Tobit analysis~\cite{Amemiya1984} is a regression method for censored data. 
In Tobit analysis, a target variable $y\in\bR$
(the concentration of a microorganism, in this study)
is assumed to be drawn with the following generative model. 
\begin{tsaligned}
      y = \left<\vw,\x\right> + \epsilon
      \end{tsaligned}
where $\epsilon$ is a normal noise, $\epsilon\sim\cN(0,\beta^{-1})$, 
the vector $\x\in\bR^{d}$ contains explanatory variables 
(including physicochemical data and possibly concentration data of another microorganism),
and $\vw\in\bR^{d}$ is a regression coefficient vector. 
This is largely the same as the setting of the least square estimation; 
however, one important difference is that Tobit analysis allows  censoring in  sample data. 
In a case where a concentration $y$ is undetected with detection limit $\theta$, the expected concentration is given by 
\begin{tsaligned}\label{eq:expect-tnl}
  \bE[y|y<\theta,\x] 
  = 
  \left<\vw,\x\right> 
  - 
  \beta^{-1/2}
  \lambda_{\text{IMR}}((\theta-\left<\vw,\x\right>)\sqrt{\beta}).
\end{tsaligned}
Herein, $\lambda_{\text{IMR}}(\xi) = \phi(\xi)/\Phi(\xi)$ is 
the \emph{inverse Mills ratio} where $\phi$ and $\Phi$ are 
the standard normal density function and its cumulative density 
function, respectively. 
Equation~\eqref{eq:expect-tnl} is derived from the fact that
under the condition $y<\theta$, 
$y$ follows the truncated normal distribution with the truncation of 
upper tail:
\begin{tsaligned}
  p(y|y<\theta,\x)
  =
  f_{\text{tn}}
  (y\,|\,\left<\vw,\x\right>,\beta,\theta)
\end{tsaligned}
where
\begin{tsaligned}
  f_{\text{tn}}
  (y\,|\,\mu,\beta,\theta)
  :=
      \begin{cases}
      \frac{\sqrt{\beta}\phi(\sqrt{\beta}(y-\mu))}
      {\Phi(\sqrt{\beta}(\theta-\mu)}
      & \text{for } y\in(-\infty,\theta), 
      \\
      0
      & \text{for } y\in [\theta,+\infty). 
   \end{cases}
\end{tsaligned}
The second moment can also be expressed in a closed form as
\begin{multline}\label{eq:secmom-tnl}
  \bE[y^{2}|y<\theta,\x]
  =
  \frac{1-\xi\lambda_{\textsc{imr}}((\theta-\left<\vw,\x\right>)\sqrt{\beta})}{\beta}
  \\
  +
  \left<\vw,\x\right>^{2}
  -
  \frac{2\lambda_{\textsc{imr}}((\theta-\left<\vw,\x\right>)\sqrt{\beta})\left<\vw,\x\right>}{\sqrt{\beta}}. 
\end{multline}

The values of the model parameters $\vw$ and $\beta$ are determined by fitting the model
to a censored dataset
$(\x_{i},y_{i})\in\bR^{d}\times\bR$ for $i=1,\dots,n$
in which $y_{1},\dots,y_{n_{\text{v}}}$ are
observed, whereas 
$y_{n_{\text{v}}+1},\dots,y_{n}$ are 
not observed due to the detection limit $\theta$. 
Fitting to the dataset is performed by 
maximizing the following regularized log-likelihood function. 
\begin{tsaligned}\label{eq:ll-symtobit}
      L_{\text{sym}}(\vw,\beta) 
      :=
      \log p_{\text{sym}}(\vw) + L_{0}(\vw,\beta),
     \end{tsaligned}      
where $p_{\text{sym}}(\vw)$ is the normal prior of 
the regression coefficients $\vw$: 
\begin{tsaligned}
      p_{\text{sym}}(\vw) = \cN(\vw\,;\,\0,\lambda^{-1}\vI). 
\end{tsaligned}      
The second term in \eqref{eq:ll-symtobit},
$L_{0}(\vw,\beta)$, 
is the Tobit log-likelihood function: 
\begin{multline}
L_{0}(\vw,\beta) 
:=
\frac{n_{\text{v}}}{2}\log\beta+
\sum_{i=1}^{n_{\text{v}}}
\log\phi
\left(\sqrt{\beta}(y_{i}-\left<\vw,\x_{i}\right>)\right)
\\
+
\sum_{i=n_{\text{v}}+1}^{n}
\log\Phi\left(\sqrt{\beta}(\theta-\left<\vw,\x_{i}\right>)\right). 
\end{multline}
The EM algorithm is a standard method for maximization of $L_{\text{sym}}$. 
The details of this method can be found in a paper by Amemiya~\cite{Amemiya1984}.

\begin{figure}[t!]
  \centering
  \footnotesize
  \begin{tabular}{lll}
    (a) $\lambda^{\text{p}}_{h}=1$,
    &
    (b) $\lambda^{\text{p}}_{h}=100$,
    &
    (c) $\lambda^{\text{p}}_{h}=1$,
    \\
    \textcolor{white}{(a)}
    $\lambda^{\text{n}}_{h}=1$,
    &
    \textcolor{white}{(b)}
    $\lambda^{\text{n}}_{h}=1$,
    &
    \textcolor{white}{(c)}
    $\lambda^{\text{n}}_{h}=100$,
    \\
    \includegraphics[width=0.15\textwidth]{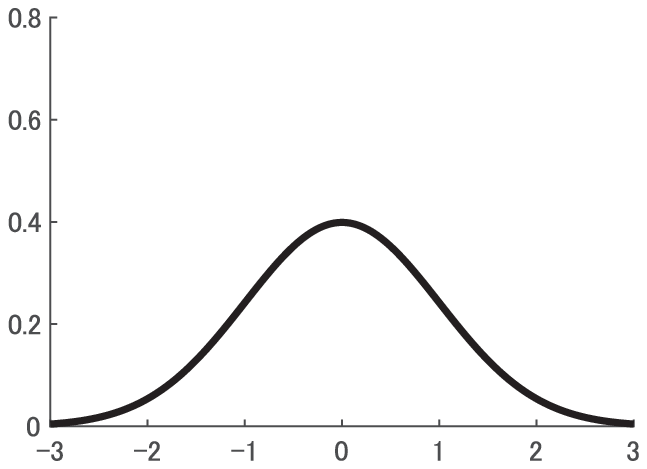}
    &
    \includegraphics[width=0.15\textwidth]{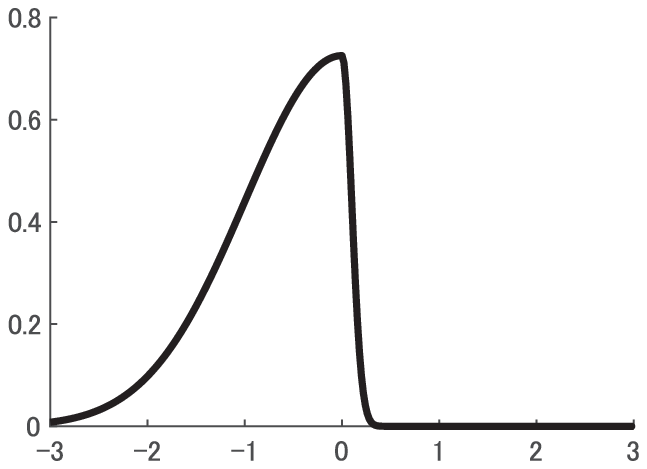}
    &
    \includegraphics[width=0.15\textwidth]{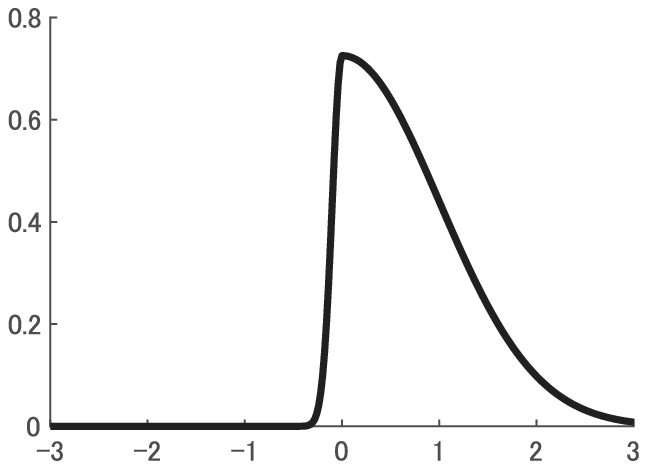}
    \\ 
  \end{tabular}
  \caption{%
  Priors of regression coefficients for asymmetric Tobit model. In the three panels, the densities $p(w_{h})$ are plotted against a regression coefficient $w_{h}$. (a) The prior is reduced to the symmetric normal distribution when $\lambda^{\text{p}}_{h}=1$ and $\lambda^{\text{n}}_{h}=1$. (b) When $\lambda^{\text{p}}_{h}\gg\lambda^{\text{n}}_{h}$, positive regression coefficients are strongly penalized. (c) When $\lambda^{\text{p}}_{h}\ll\lambda^{\text{n}}_{h}$, negative coefficients are likely to be avoided. %
  \label{fig:test904}
  }
\end{figure}

\subsection{Nonnegative least square}
\label{ss:nnls}
The nonnegative least square problem is a quadratic programming problem defined as 
\begin{tsaligned}
  \text{min}\quad
  &
  \lVert \vA^\top\x-\vb \rVert
  \quad
  \text{wrt}\quad
  \x\in\bR^{m},
  \\
  \text{where}\quad
  &
  \vA\in\bR^{m\times n}, \vb\in\bR^{n}. 
\end{tsaligned}
This problem is denoted by 
$\text{NNLS}(\vA,\vb)$ hereinafter.  
For solving the NNLS problem, Lawson and Hanson's active set algorithm presented in their book~\cite{Lawson1995solving} is popular. Since then, many improvements have been developed, and presently, NNLS is known as an efficiently solvable convex problem~\cite{DonghuiChen09-nonneg,Meinshausen2013,Bellavia2006}.

\section{Correlation analysis methods}
\label{s:pccest}
In this study, we consider three approaches for correlation analysis: a na\"{i}ve approach, a classical Tobit approach, and an asymmetric Tobit approach. The three approaches are summarized in Figure~\ref{fig:3meth}. The details are described below. 

\textbf{Na\"{i}ve approach:} 
Assume that a dataset contains $n$ data pairs
\tslong{%
\begin{tsaligned}
  (y_{1,\text{a}},y_{1,\text{b}}),\dots,(y_{n,\text{a}},y_{n,\text{b}}) 
\end{tsaligned}
}
\tsshort{%
\(
  (y_{1,\text{a}},y_{1,\text{b}}),\dots,(y_{n,\text{a}},y_{n,\text{b}}) 
\)
}
representing concentrations of two microorganisms that may be left-censored. 
Let $\theta_{\text{a}}$ and $\theta_{\text{b}}$ be the detection limits of the two microorganisms, respectively.  
The data are such that $y_{i,\text{a}}<\theta_{\text{a}}$ and $y_{i,\text{b}}<\theta_{\text{b}}$ are not available. 
\tslong{%
We use the index sets of visible entries 
\begin{tsaligned}
  &\cI_{\text{v,a}} := \left\{ i\in[n]\,\middle|\,y_{i,\text{a}} \ge \theta_{\text{a}}\right\}
  \quad\text{and}
  \\
  &
  \cI_{\text{v,b}} := \left\{ i\in[n]\,\middle|\,y_{i,\text{b}} \ge \theta_{\text{b}} \right\}. 
\end{tsaligned}
Our example of a na\"{i}ve method computes PCC only from visible pairs 
(i.e. $(y_{i,\text{a}},y_{i,\text{b}})$ for $i\in\cI_{\text{vv}}:=\cI_{\text{v,a}}\cap\cI_{\text{v,b}}$). 
Namely, the PCC is computed as
\begin{tsaligned}
  R_{\text{na\"{i}ve}}
  :=
  R(\y_{\text{vv,a}},\y_{\text{vv,b}})
\end{tsaligned}      
where
\begin{tsaligned}
  & \y_{\text{vv,a}} 
  := 
  \left[ y_{i,\text{a}} \right]_{i\in\cI_{\text{vv}}}, 
  \quad
  \y_{\text{vv,b}} 
  := 
  \left[ y_{i,\text{b}} \right]_{i\in\cI_{\text{vv}}}. 
\end{tsaligned}      
}
\tsshort{
  A na\"{i}ve method computes PCC only from visible pairs. We denote by $R_{\text{na\"{i}ve}}$ the PCC computed with the na\"{i}ve method. 
}
A shortcoming of this approach is that the cardinality of commonly visible set $\cI_{\text{vv}}$ tends to be small, yielding a large estimation error.  

\textbf{Classical Tobit approach:} 
We now consider another approach to correlation analysis utilizing undetected entries of the concentrations of two microorganisms A and B. Here, it is assumed that other physicochemical observations are available as side information. Typical physicochemical data such as water temperature, DO, SS, TN, and TP are more easily measured compared to microorganism concentrations. The approach being discussed here imputes undetected concentrations of the microorganism B, and then imputes the undetected concentrations of the microorganism A using the side information and B's completed concentrations. Tobit analysis is used for imputation of undetected concentrations. This method is referred to as the \emph{classical Tobit approach}. 

After the above procedure, the concentration data of both microorganisms are complete. 
PCC can be computed from the completed vectors as
\tslong{%
\begin{tsaligned}
  R_{\text{sym}}
  :=
  R(\hat{\y}_{\text{a}},\hat{\y}_{\text{b}}), 
\end{tsaligned}      
}
\tsshort{%
\(
  R_{\text{sym}}
  :=
  R(\hat{\y}_{\text{a}},\hat{\y}_{\text{b}}), 
\)
}
where the completed vectors are denoted by 
$\hat{\y}_{\text{a}} := \left[ y_{i,\text{a}} \right]_{i\in[n]}$ 
and 
$\hat{\y}_{\text{b}} := \left[ y_{i,\text{b}} \right]_{i\in[n]}$, 
respectively. 
PCC is expected to be estimated well if the imputations of undetected entries are accurate. 

\textbf{Asymmetric Tobit approach:} 
The third approach exploits domain information to improve the Tobit analysis, and consequently, the PCC estimation. In water quality engineering, it is known whether typical physicochemical data are positively correlated to each of several typical pathogens. For example, more pathogens tend to survive in warmer water, leading to positive correlation between pathogen concentration and water temperature. It can be assumed that all correlated explanatory variables have positive correlations to a target variable without loss of generality, because negatively correlated explanatory variables are negated in advance by preprocessing. 
For positively correlated explanatory variables, positive regression coefficients are preferred. 
However, in highly censored datasets, often only a few visible observations are available. 
In such a case, positively correlated explanatory variables may often have a negative sample correlation in small samples, which decreases the effectiveness of the Tobit model.  
The third approach to correlation analysis uses a modification of the Tobit model, introduced below, to impute undetected concentrations. We denote the resultant PCC by $R_{\text{asym}}$. 

In the rest of this section, our proposed modification of the Tobit model is described. 
This modified Tobit model is called the \emph{asymmetric Tobit model}, and 
the correlation analysis approach using the new Tobit model is called 
the \emph{asymmetric Tobit approach} hereinafter. 
Asymmetric Tobit model penalizes the negative coefficient. 
To do so, the ordinary normal prior in \eqref{eq:ll-symtobit} is replaced by the asymmetric normal distribution~\cite{KatOmaAso02a} (See Figure~\ref{fig:test904}) as follows. 
\begin{tsaligned}
  p_{\text{asym}}(\vw) := 
  \prod_{h=1}^{d}
  \frac{1}{Z_{h}}
  \exp
  \left(
  -\frac
  {\lambda^{\text{p}}_{h} (w_{h})_{+}^{2}
  +\lambda^{\text{n}}_{h} (-w_{h})_{+}^{2}}{2}
  \right)
\end{tsaligned}
where $(x)_{+}:=\max(0,x)$ and
\begin{tsaligned}
  Z_{h} := 
  \sqrt{\frac{\pi}{ 2\lambda^{\text{p}}_{h}}} 
  +
  \sqrt{\frac{\pi}{ 2\lambda^{\text{n}}_{h}}}. 
\end{tsaligned}
Let $\cI_{\text{p}} \subseteq [d]$ be the index set of explanatory variables correlated to the target variable. 
In our simulations described later, 
the constant vectors $\vlam^{\text{p}}, \vlam^{\text{h}}\in\bR^{d}$ 
are set to 
$\lambda^{\text{p}}_{h}=  (1 + 99\mathds{1}[h\in\cI_{\text{n}}])\lambda$ and 
$\lambda^{\text{n}}_{h}=  (1 + 99\mathds{1}[h\in\cI_{\text{p}}])\lambda$ for $h\in[d]$. 
The new regularized log-likelihood function is expressed as
\begin{tsaligned}\label{eq:ll-asym}
L_{\text{asym}}(\vw,\beta) 
:=
\log p_{\text{asym}}(\vw) + L_{0}(\vw,\beta). 
\end{tsaligned}      
The new Tobit model is fitted to censored data by maximizing the new regularized log-likelihood function~\eqref{eq:ll-asym}. 
In the next section, our approach to maximizing the new objective function~\eqref{eq:ll-asym} is described.

\section{Fitting asymmetric Tobit model}
\label{s:asym}
In this study, we propose a new algorithm for fitting the asymmetric Tobit model. To find the maximizer of the regularized log-likelihood function~\eqref{eq:ll-asym}, we adopted the expectation-maximization (EM) algorithm. 
Modification of the prior often gives rise to 
some technical difficulties. 
In this section, we show that 
each iteration of the EM algorithm
can be performed efficiently even if 
the prior is changed from the ordinary normal distribution to the asymmetric normal distribution. 

EM algorithms are a general framework for fitting a latent variable model to a dataset by repeating E-step and M-step until convergence. 
The EM algorithm for Tobit analysis uses the following Q-function. 
\begin{tsaligned}
    &Q(\vw,\beta,q)
    :=
    \log p(\vw) + 
    \frac{n}{2}\log\beta
    \\
    &+
    \sum_{i=1}^{n_{\text{v}}}
    \log\phi
    \left(\sqrt{\beta}(y_{i}-\left<\vw,\x_{i}\right>)\right)
    \\
    &+
    \sum_{i=n_{\text{v}}+1}^{n}
    \bE_{q_{i}(y_{i})}
    \left[
    \log\phi
    \left(\sqrt{\beta}(y_{i}-\left<\vw,\x_{i}\right>)\right)
    \right]
\end{tsaligned}
where $q$ is a set of $(n-n_{\text{v}}$) probabilistic density functions
$q_{n_{\text{v}}+1}(y_{n_{\text{v}}+1}),\dots,q_{n}(y_{n})$. 
Therein, $p(\vw)$ is the prior of $\vw$; $p=p_{\text{sym}}$ for 
the classical Tobit model and $p=p_{\text{asym}}$ for 
the asymmetric Tobit model. 
Let $(\vw^{(t-1)},\beta^{(t-1)})$ denote the value 
of the model parameters obtained at $(t-1)$th iteration. 
The set of the distributions $q$ at the $t$-th iteration is denoted by 
$q^{(t)}:=\left(q_{i}^{(t)}\right)_{i=n_{\text{v}}+1}^{n}$. 
The $t$th iteration consists of the following procedure. 
\begin{enumerate}
\item Set the density function~$q_{i}^{(t)}$ to the posterior of $y_{i}$ based on $(\vw^{(t-1)},\beta^{(t-1)})$, 
and update each of the expected terms in the Q-function. 
\item $\vw^{(t)} := \argmax_{\vw\in\bR^{d}}Q(\vw,\beta^{(t-1)},q^{(t)})$; 
\item $\beta^{(t)} := \argmax_{\beta\in\bR}Q(\vw^{(t)},\beta,q^{(t)})$; 
\end{enumerate}
The first line is called the E-step. The other two lines are 
called the M-step. The E-step and the update rule of $\beta$ are 
unchanged even if the prior of $\vw$ is changed. Meanwhile, the 
change of the prior of $\vw$ may complicate the update rule of $\vw$. 
In this study, we found the following result. 
\begin{theorem-waku}\label{thm:asym-mstep-nnls}
    If $p=p_{\text{asym}}$, the update rule of $\vw$ in the EM algorithm
    for fitting the Tobit model is reduced to an NNLS problem. 
\end{theorem-waku}
This theorem implies that each iteration of the EM algorithm is performed efficiently even if the prior of the regression coefficients $\vw$ is replaced with the asymmetric normal distribution.

Before discussing the update rule of $\vw$, we review the E-step and the update rule of $\beta$. 
Let 
\begin{align*}
    &\y^{\text{v}}:=\left[ y_{1},\dots,y_{n_{\text{v}}}\right]^\top, \quad
    &&
    \y^{\text{h}}:=\left[ y_{n_{\text{v}}+1},\dots,y_{n}\right]^\top, 
    \\
    &\X^{\text{v}}:=\left[ \x_{1},\dots,\x_{n_{\text{v}}}\right], \quad
    &&
    \X^{\text{h}}:=\left[ \x_{n_{\text{v}}+1},\dots,\x_{n}\right].  
\end{align*}
The posterior, computed at the E-step of $t$th iteration, is updated as 
\begin{tsaligned}\label{eq:qt-update}
    q_{i}^{(t)}(y_{i}) = 
    f_{\text{tn}}
    \left(y_{i}\,\middle|\,\left<\vw^{(t-1)},\x_{i}\right>,\beta^{(t-1)},\theta\right). 
\end{tsaligned}
This allows us to update the following expected quantities. 
\begin{tsaligned}\label{eq:baryt-vt-def}
    &\bar{\y}^{(t)} :=
    \left[
      \left(\y^{\rv}\right)^\top, \,
      \bE_{q^{(t)}}\left[\left(\y^{\rh}\right)^\top\right]
      \right]^\top,
    \\
    &v^{(t)}:=
    \bE_{q^{(t)}}\left[\left\lVert\y^{\rh}\right\rVert^{2}\right]
    -
    \left\lVert\bE_{q^{(t)}}\left[\y^{\rh}\right]\right\rVert^{2} . 
\end{tsaligned}
Each expectation
in both $\bar{\y}^{(t)}$ and $v^{(t)}$
is expressed in a closed form using 
\eqref{eq:expect-tnl} and \eqref{eq:secmom-tnl}. 
The update rule of $\beta$ is readily obtained
by setting the derivative of the Q-function as
\begin{tsaligned}\label{eq:beta-update-closed-form}
    \beta^{(t)}
    =
    \frac{n}{%
    \lVert\X^\top\vw-\bar{\y}^{(t)}\rVert^{2}
    +
    v^{(t)}}. 
\end{tsaligned}
We thus observe that efficient computation
of the E-step and the update rule of $\beta$ is possible. 

Finally, 
we conclude this section by demonstrating that NNLS fitting accomplishes the update rule of $\vw$,
as described in Theorem~\ref{thm:asym-mstep-nnls}. 
Define a $2d\times(n+2d)$ matrix $\vA^{(t)}$
and an $(n+2d)$-dimensional vector $\vb^{(t)}$ as
\begin{tsaligned}\label{eq:At-bt-def-in-em}
&\vA^{(t)}
:=
\begin{bmatrix}
    \X & 
    \diag\left(
        \frac{\vlam^{\text{p}}}{\beta^{(t-1)}}\right)^{1/2} & \vO
    \\
    -\X & \vO & 
    \text{diag}\left(\frac{\vlam^{\text{n}}}{\beta^{(t-1)}}\right)^{1/2}
\end{bmatrix}, 
\\
&\text{and}
\quad
\vb^{(t)}
:=
\begin{bmatrix}
    \bar{\y}^{(t)} \\ \0_{2d}
\end{bmatrix}. 
\end{tsaligned}
The regression coefficient vector $\vw\in\bR^{d}$ can be decomposed with two nonnegative vectors $\vw_{+},\vw_{-}\in\bR_{+}^{d}$ as $\vw=\vw_{+}-\vw_{-}$. Using the two vectors, the Q-function can be rewritten as
\begin{multline}\label{eq:Q-for-w-update}
    Q(\vw_{+}-\vw_{-},\beta^{(t-1)},q^{(t)})
=
\\
-\frac{\beta}{2}
\left\lVert
(\vA^{(t)})^\top
\begin{bmatrix} \vw_{+} \\ \vw_{-} \end{bmatrix}
-
\vb^{(t)}
\right\rVert^{2}
 + \text{const}
\end{multline}
where $\text{const}$ denotes the terms with no dependency on the regression coefficients. 
Equation~\eqref{eq:Q-for-w-update} implies that the sub-problem for maximizing 
$Q(\cdot,\beta^{(t-1)},q^{(t)})$ 
is reduced to the problem 
$\text{NNLS}(\vA^{(t)},\vb^{(t)})$
defined in Subsection~\ref{ss:nnls}. 
From the optimal solution to the sub-problem, 
denoted by 
$\begin{bmatrix} \vw_{+}^{(t)} \\ \vw_{-}^{(t)}\end{bmatrix}$, 
the regression coefficient vector is updated as
$\vw^{(t)}=\vw_{+}^{(t)}-\vw_{-}^{(t)}$. 

The above discussions are summarized in Algorithm~\ref{algo:em-for-tobit}
that shows a pseudo-code of the EM algorithm for fitting 
the asymmetric Tobit model. 
\begin{algorithm}[t!]
    \caption{
        EM algorithm for asymmetric Tobit model. 
        \label{algo:em-for-tobit}}
    \Begin{
        Initialize $\vw^{(0)}$ and $\beta^{(0)}$\; 
        \For{$t:=1$ \KwTo $T$}{
            Use \eqref{eq:qt-update} and \eqref{eq:baryt-vt-def} to update $q$ and compute $\bar{\y}^{(t)}$ and $v^{(t)}$\; 
            Solve $\text{NNLS}(\vA^{(t)},\vb^{(t)})$ where $\vA^{(t)}$ and $\vb^{(t)}$ are defined as \eqref{eq:At-bt-def-in-em} to get $\vw_{+}^{(t)}$ and $\vw_{-}^{(t)}$\; 
            $\vw^{(t)}:=\vw_{+}^{(t)}-\vw_{-}^{(t)}$\;
            Update the inverse variance parameter by \eqref{eq:beta-update-closed-form}\;
        }
    }
\end{algorithm}

\begin{table}[t]
    \caption{Estimation errors on Indian water dataset.}
    \centering
    \footnotesize
    \begin{tabular}{|c|c||c|c|c|}
    \hline
    A & B & Asym Tobit & Sym Tobit & Na\"{i}ve 
    \\
    \hline
    FC & TC &  {\textbf{ {0.025}}} (0.020)  &  {0.075} (0.064)  &  {0.083} (0.100) \\
    FC & pH &  {\textbf{ {0.134}}} (0.104)  &  {0.171} (0.115)  &  {0.623} (0.267) \\
    FC & Cond &  {\textbf{ {0.112}}} (0.091)  &  { {0.119}} (0.093)  &  {0.522} (0.335) \\
    FC & N &  {\textbf{ {0.116}}} (0.083)  &  {0.131} (0.092)  &  {0.419} (0.272) \\
    FC & BOD &  { {0.156}} (0.123)  &  {\textbf{ {0.151}}} (0.132)  &  {0.453} (0.250) \\
    TC & FC &  {\textbf{ {0.028}}} (0.022)  &  {0.060} (0.057)  &  {0.083} (0.100) \\
    TC & pH &  {\textbf{ {0.116}}} (0.084)  &  {0.163} (0.115)  &  {0.635} (0.308) \\
    TC & Cond &  {\textbf{ {0.142}}} (0.082)  &  {0.178} (0.104)  &  {0.732} (0.343) \\
    TC & N &  {\textbf{ {0.101}}} (0.081)  &  {0.114} (0.087)  &  {0.376} (0.313) \\
    TC & BOD &  {\textbf{ {0.091}}} (0.069)  &  { {0.096}} (0.068)  &  {0.441} (0.347) \\
    pH & FC &  {\textbf{ {0.141}}} (0.091)  &  {0.191} (0.107)  &  {0.623} (0.267) \\
    pH & TC &  {\textbf{ {0.124}}} (0.091)  &  {0.165} (0.116)  &  {0.635} (0.308) \\
    pH & Cond &  {\textbf{ {0.144}}} (0.068)  &  {0.167} (0.081)  &  {0.684} (0.364) \\
    pH & N &  {\textbf{ {0.114}}} (0.094)  &  {0.127} (0.107)  &  {0.978} (0.036) \\
    pH & BOD &  {\textbf{ {0.131}}} (0.087)  &  { {0.156}} (0.123)  &  {0.600} (0.318) \\
    Cond & FC &  {\textbf{ {0.098}}} (0.081)  &  { {0.111}} (0.086)  &  {0.522} (0.335) \\
    Cond & TC &  {\textbf{ {0.135}}} (0.078)  &  {0.167} (0.093)  &  {0.729} (0.341) \\
    Cond & pH &  {\textbf{ {0.128}}} (0.068)  &  {0.161} (0.081)  &  {0.684} (0.364) \\
    Cond & N &  {\textbf{ {0.066}}} (0.060)  &  {0.090} (0.074)  &  {0.558} (0.302) \\
    Cond & BOD &  {\textbf{ {0.066}}} (0.049)  &  { {0.070}} (0.051)  &  {0.518} (0.318) \\
    N & FC &  {\textbf{ {0.133}}} (0.087)  &  {0.145} (0.093)  &  {0.419} (0.272) \\
    N & TC &  {\textbf{ {0.115}}} (0.088)  &  { {0.127}} (0.098)  &  {0.376} (0.313) \\
    N & pH &  {\textbf{ {0.070}}} (0.052)  &  {0.098} (0.080)  &  {0.978} (0.036) \\
    N & Cond &  {\textbf{ {0.071}}} (0.064)  &  {0.098} (0.080)  &  {0.558} (0.302) \\
    N & BOD &  {\textbf{ {0.143}}} (0.072)  &  {\textbf{ {0.143}}} (0.072)  &  {0.342} (0.215) \\
    BOD & FC &  { {0.165}} (0.118)  &  {\textbf{ {0.161}}} (0.140)  &  {0.453} (0.250) \\
    BOD & TC &  {\textbf{ {0.103}}} (0.075)  &  { {0.110}} (0.078)  &  {0.441} (0.347) \\
    BOD & pH &  {\textbf{ {0.111}}} (0.088)  &  {0.154} (0.130)  &  {0.600} (0.318) \\
    BOD & Cond &  {\textbf{ {0.053}}} (0.041)  &  { {0.070}} (0.057)  &  {0.518} (0.318) \\
    BOD & N &  {\textbf{ {0.139}}} (0.085)  &  { {0.143}} (0.088)  &  {0.342} (0.215) \\
        \hline
    \end{tabular}
    \label{tab:indian}
\end{table}
\tslong{%
\begin{table}[t]
    \caption{Estimation errors on Harbor water dataset.}
    \centering    
    \footnotesize
    \begin{tabular}{|c|c||c|c|c|}
    \hline
    A & B & Asym Tobit & Sym Tobit & Na\"{i}ve 
    \\
    \hline
    FC & TC &  {\textbf{ {0.102}}} (0.066)  &  { {0.109}} (0.074)  &  {0.475} (0.429) \\
    FC & WT &  {\textbf{ {0.138}}} (0.099)  &  { {0.149}} (0.100)  &  {0.647} (0.279) \\
    FC & pH &  {\textbf{ {0.197}}} (0.127)  &  {\textbf{ {0.197}}} (0.124)  &  {0.567} (0.298) \\
    TC & FC &  {\textbf{ {0.129}}} (0.078)  &  { {0.139}} (0.084)  &  {0.475} (0.429) \\
    TC & WT &  {\textbf{ {0.136}}} (0.108)  &  {0.151} (0.116)  &  {0.656} (0.368) \\
    TC & pH &  {\textbf{ {0.102}}} (0.066)  &  { {0.109}} (0.074)  &  {0.475} (0.429) \\
    WT & FC &  {\textbf{ {0.147}}} (0.101)  &  { {0.153}} (0.110)  &  {0.647} (0.279) \\
    WT & TC &  {\textbf{ {0.149}}} (0.107)  &  { {0.157}} (0.116)  &  {0.656} (0.368) \\
    WT & pH &  {\textbf{ {0.149}}} (0.107)  &  { {0.157}} (0.116)  &  {0.656} (0.368) \\
    pH & FC &  {\textbf{ {0.195}}} (0.128)  &  {\textbf{ {0.195}}} (0.124)  &  {0.567} (0.298) \\
    pH & TC &  {\textbf{ {0.129}}} (0.078)  &  { {0.139}} (0.084)  &  {0.475} (0.429) \\
    pH & WT &  {\textbf{ {0.136}}} (0.108)  &  {0.151} (0.116)  &  {0.656} (0.368) \\
        \hline
    \end{tabular}
    \label{tab:harbor}
\end{table}
}
\tslong{%
\begin{table*}[t]
    \caption{Estimation errors on Sapporo water dataset.}
    \centering    
    \footnotesize
    \begin{tabular}{cc}
    \begin{tabular}{|c|c||c|c|c|}
    \hline
    A & B & Asym Tobit & Sym Tobit & Na\"{i}ve 
    \\
    \hline
    E.coli & TC &  {0.110} (0.027)  &  {\textbf{ {0.082}}} (0.039)  &  {0.776} (0.268) \\
    E.coli & pH &  {\textbf{ {0.087}}} (0.068)  &  {0.094} (0.066)  &  {0.337} (0.232) \\
    E.coli & EC &  {\textbf{ {0.116}}} (0.082)  &  {0.132} (0.089)  &  {0.993} (0.541) \\
    E.coli & SS &  {\textbf{ {0.115}}} (0.089)  &  { {0.133}} (0.102)  &  {0.631} (0.390) \\
    E.coli & TN &  {0.101} (0.057)  &  {\textbf{ {0.059}}} (0.048)  &  {0.651} (0.329) \\
    E.coli & TP &  {0.180} (0.053)  &  {\textbf{ {0.138}}} (0.063)  &  {0.731} (0.492) \\
    E.coli & FR &  {\textbf{ {0.264}}} (0.117)  &  { {0.270}} (0.117)  &  {1.058} (0.288) \\
    TC & E.coli &  {0.103} (0.031)  &  {\textbf{ {0.085}}} (0.033)  &  {0.776} (0.268) \\
    TC & pH &  { {0.127}} (0.081)  &  {\textbf{ {0.124}}} (0.081)  &  {0.574} (0.380) \\
    TC & EC &  {\textbf{ {0.071}}} (0.048)  &  { {0.072}} (0.045)  &  {0.654} (0.458) \\
    TC & SS &  {\textbf{ {0.112}}} (0.079)  &  { {0.121}} (0.096)  &  {0.333} (0.152) \\
    TC & TN &  {0.105} (0.055)  &  {\textbf{ {0.069}}} (0.046)  &  {0.917} (0.445) \\
    TC & TP &  {0.171} (0.050)  &  {\textbf{ {0.109}}} (0.060)  &  {0.764} (0.496) \\
    TC & FR &  {\textbf{ {0.167}}} (0.081)  &  {0.206} (0.078)  &  {0.563} (0.305) \\
    pH & E.coli &  {\textbf{ {0.098}}} (0.080)  &  {0.111} (0.082)  &  {0.337} (0.232) \\
    pH & TC &  {\textbf{ {0.118}}} (0.077)  &  { {0.120}} (0.085)  &  {0.574} (0.380) \\
    pH & EC &  {\textbf{ {0.082}}} (0.054)  &  {\textbf{ {0.082}}} (0.054)  &  {0.451} (0.314) \\
    pH & SS &  {\textbf{ {0.095}}} (0.071)  &  { {0.098}} (0.078)  &  {0.675} (0.343) \\
    pH & TN &  {\textbf{ {0.179}}} (0.114)  &  {0.191} (0.119)  &  {0.962} (0.354) \\
    pH & TP &  { {0.121}} (0.100)  &  {\textbf{ {0.111}}} (0.093)  &  {0.483} (0.330) \\
    pH & FR &  {\textbf{ {0.138}}} (0.104)  &  {0.166} (0.120)  &  {0.734} (0.292) \\
    EC & E.coli &  {\textbf{ {0.114}}} (0.084)  &  {0.129} (0.096)  &  {0.998} (0.542) \\
    EC & TC &  { {0.086}} (0.062)  &  {\textbf{ {0.084}}} (0.057)  &  {0.654} (0.458) \\
    EC & pH &  { {0.078}} (0.061)  &  {\textbf{ {0.077}}} (0.061)  &  {0.451} (0.314) \\
    EC & SS &  {\textbf{ {0.082}}} (0.065)  &  {0.125} (0.095)  &  {0.767} (0.348) \\
    EC & TN &  {\textbf{ {0.232}}} (0.092)  &  {0.244} (0.094)  &  {0.502} (0.381) \\
    EC & TP &  {\textbf{ {0.067}}} (0.045)  &  {0.078} (0.046)  &  {0.620} (0.390) \\
    EC & FR &  {\textbf{ {0.247}}} (0.076)  &  {0.297} (0.106)  &  {1.062} (0.214) \\
    \hline
\end{tabular}
&
\begin{tabular}{|c|c||c|c|c|}
    \hline
    SS & E.coli &  {\textbf{ {0.122}}} (0.090)  &  { {0.139}} (0.105)  &  {0.631} (0.390) \\
    SS & TC &  {\textbf{ {0.118}}} (0.087)  &  { {0.136}} (0.103)  &  {0.333} (0.152) \\
    SS & pH &  {\textbf{ {0.095}}} (0.074)  &  {\textbf{ {0.095}}} (0.081)  &  {0.675} (0.343) \\
    SS & EC &  {\textbf{ {0.080}}} (0.059)  &  {0.101} (0.086)  &  {0.767} (0.348) \\
    SS & TN &  {\textbf{ {0.163}}} (0.114)  &  {0.188} (0.129)  &  {0.319} (0.221) \\
    SS & TP &  {\textbf{ {0.146}}} (0.103)  &  {0.174} (0.132)  &  {1.049} (0.169) \\
    SS & FR &  {\textbf{ {0.086}}} (0.060)  &  {0.116} (0.084)  &  {0.495} (0.290) \\
    TN & E.coli &  {0.120} (0.056)  &  {\textbf{ {0.078}}} (0.056)  &  {0.651} (0.329) \\
    TN & TC &  {0.130} (0.043)  &  {\textbf{ {0.078}}} (0.051)  &  {0.925} (0.448) \\
    TN & pH &  {\textbf{ {0.125}}} (0.091)  &  { {0.131}} (0.095)  &  {0.962} (0.354) \\
    TN & EC &  {\textbf{ {0.229}}} (0.089)  &  {0.240} (0.090)  &  {0.502} (0.381) \\
    TN & SS &  {\textbf{ {0.145}}} (0.111)  &  {0.190} (0.136)  &  {0.316} (0.219) \\
    TN & TP &  {0.055} (0.036)  &  {\textbf{ {0.046}}} (0.030)  &  {0.716} (0.286) \\
    TN & FR &  {\textbf{ {0.224}}} (0.099)  &  {0.264} (0.108)  &  {0.524} (0.298) \\
    TP & E.coli &  {0.181} (0.054)  &  {\textbf{ {0.139}}} (0.063)  &  {0.731} (0.492) \\
    TP & TC &  {0.176} (0.046)  &  {\textbf{ {0.128}}} (0.057)  &  {0.756} (0.494) \\
    TP & pH &  { {0.133}} (0.114)  &  {\textbf{ {0.125}}} (0.105)  &  {0.483} (0.330) \\
    TP & EC &  {\textbf{ {0.082}}} (0.044)  &  {0.094} (0.045)  &  {0.620} (0.390) \\
    TP & SS &  {\textbf{ {0.122}}} (0.089)  &  { {0.149}} (0.115)  &  {1.052} (0.179) \\
    TP & TN &  { {0.054}} (0.030)  &  {\textbf{ {0.046}}} (0.028)  &  {0.716} (0.286) \\
    TP & FR &  {\textbf{ {0.153}}} (0.113)  &  {0.207} (0.128)  &  {0.436} (0.269) \\
    FR & E.coli &  {\textbf{ {0.260}}} (0.118)  &  { {0.271}} (0.117)  &  {1.058} (0.288) \\
    FR & TC &  {\textbf{ {0.208}}} (0.111)  &  {0.253} (0.112)  &  {0.563} (0.305) \\
    FR & pH &  {\textbf{ {0.129}}} (0.105)  &  {0.145} (0.109)  &  {0.734} (0.292) \\
    FR & EC &  {\textbf{ {0.297}}} (0.086)  &  {0.326} (0.104)  &  {1.062} (0.214) \\
    FR & SS &  {\textbf{ {0.083}}} (0.059)  &  {0.109} (0.082)  &  {0.487} (0.290) \\
    FR & TN &  {\textbf{ {0.259}}} (0.096)  &  {0.285} (0.096)  &  {0.524} (0.298) \\
    FR & TP &  {\textbf{ {0.240}}} (0.158)  &  {0.300} (0.169)  &  {0.421} (0.264) \\
    \hline
\end{tabular}
\end{tabular}
\label{tab:sapporo}
\end{table*}
}
\tslong{%
\newcommand{\tsunderbarcomptime}[1]{{#1}}

\begin{table*}[t]
    \caption{Computational times on four datasets.}
    \centering    
    \footnotesize
    \begin{tabular}[t]{p{0.3\linewidth} p{0.3\linewidth} p{0.3\linewidth}}
        (a) Indian & $\quad$ (b) NY Harbor & (d) Random
        \\
        \begin{tabular}[t]{|c|c|c|}
            \hline
            $n$ & Asym Tobit & Sym Tobit\\
           \hline
           10 & \textcolor{black}{0.330} (0.001)  & \textcolor{black}{\textbf{\tsunderbarcomptime{0.321}}} (0.001) \\
           17 & \textcolor{black}{0.536} (0.001)  & \textcolor{black}{\textbf{\tsunderbarcomptime{0.529}}} (0.001) \\
           31 & \textcolor{black}{\textbf{\tsunderbarcomptime{0.957}}} (0.006)  & \textcolor{black}{\tsunderbarcomptime{0.964}} (0.012) \\
           56 & \textcolor{black}{\textbf{\tsunderbarcomptime{1.706}}} (0.002)  & \textcolor{black}{1.715} (0.003) \\
           100 & \textcolor{black}{\tsunderbarcomptime{3.026}} (0.010)  & \textcolor{black}{\textbf{\tsunderbarcomptime{3.025}}} (0.002) \\
           177 & \textcolor{black}{\textbf{\tsunderbarcomptime{5.315}}} (0.011)  & \textcolor{black}{\tsunderbarcomptime{5.322}} (0.001) \\
           316 & \textcolor{black}{\textbf{\tsunderbarcomptime{9.429}}} (0.034)  & \textcolor{black}{\tsunderbarcomptime{9.434}} (0.032) \\
           562 & \textcolor{black}{\tsunderbarcomptime{16.665}} (0.046)  & \textcolor{black}{\textbf{\tsunderbarcomptime{16.633}}} (0.021) \\
           1000 & \textcolor{black}{\textbf{\tsunderbarcomptime{29.610}}} (0.094)  & \textcolor{black}{\tsunderbarcomptime{29.660}} (0.080) \\
           \hline
        \end{tabular}
        &
        \begin{tabular}[t]{l}
        \begin{tabular}[t]{|c|c|c|}
            \hline
           $n$ & Asym Tobit & Sym Tobit\\
           \hline
           10 & \textcolor{black}{0.332} (0.002)  & \textcolor{black}{\textbf{\tsunderbarcomptime{0.324}}} (0.001) \\
           17 & \textcolor{black}{0.541} (0.002)  & \textcolor{black}{\textbf{\tsunderbarcomptime{0.535}}} (0.001) \\
           31 & \textcolor{black}{0.959} (0.002)  & \textcolor{black}{\textbf{\tsunderbarcomptime{0.952}}} (0.001) \\
           56 & \textcolor{black}{\tsunderbarcomptime{1.703}} (0.002)  & \textcolor{black}{\textbf{\tsunderbarcomptime{1.701}}} (0.007) \\
           100 & \textcolor{black}{\textbf{\tsunderbarcomptime{2.986}}} (0.007)  & \textcolor{black}{\tsunderbarcomptime{2.990}} (0.012) \\
           177 & \textcolor{black}{\textbf{\tsunderbarcomptime{5.276}}} (0.023)  & \textcolor{black}{\tsunderbarcomptime{5.280}} (0.011) \\
           \hline
           \end{tabular}        
        \\ \\
        (c) Sapporo  
        \\
        \begin{tabular}[t]{|c|c|c|}
            \hline
           $n$ & Asym Tobit & Sym Tobit\\
           \hline
           10 & \textcolor{black}{0.335} (0.001)  & \textcolor{black}{\textbf{\tsunderbarcomptime{0.326}}} (0.000) \\
           17 & \textcolor{black}{0.544} (0.001)  & \textcolor{black}{\textbf{\tsunderbarcomptime{0.535}}} (0.001) \\
           31 & \textcolor{black}{0.962} (0.001)  & \textcolor{black}{\textbf{\tsunderbarcomptime{0.956}}} (0.001) \\
           56 & \textcolor{black}{1.709} (0.002)  & \textcolor{black}{\textbf{\tsunderbarcomptime{1.700}}} (0.002) \\
           100 & \textcolor{black}{\tsunderbarcomptime{3.019}} (0.002)  & \textcolor{black}{\textbf{\tsunderbarcomptime{3.011}}} (0.005) \\
           \hline
        \end{tabular}
        \end{tabular}
        &
        \begin{tabular}[t]{|c|c|c|}
            \hline
           $n$ & Asym Tobit & Sym Tobit\\
           \hline
        10 & \textcolor{black}{1.028} (0.070)  & \textcolor{black}{\textbf{\tsunderbarcomptime{0.342}}} (0.005) \\
        17 & \textcolor{black}{1.331} (0.117)  & \textcolor{black}{\textbf{\tsunderbarcomptime{0.549}}} (0.002) \\
        31 & \textcolor{black}{1.804} (0.095)  & \textcolor{black}{\textbf{\tsunderbarcomptime{0.969}}} (0.006) \\
        56 & \textcolor{black}{3.041} (0.099)  & \textcolor{black}{\textbf{\tsunderbarcomptime{1.720}}} (0.007) \\
        100 & \textcolor{black}{5.226} (0.248)  & \textcolor{black}{\textbf{\tsunderbarcomptime{3.032}}} (0.017) \\
        177 & \textcolor{black}{6.367} (0.275)  & \textcolor{black}{\textbf{\tsunderbarcomptime{5.313}}} (0.018) \\
        316 & \textcolor{black}{10.626} (0.282)  & \textcolor{black}{\textbf{\tsunderbarcomptime{9.414}}} (0.033) \\
        562 & \textcolor{black}{18.297} (0.121)  & \textcolor{black}{\textbf{\tsunderbarcomptime{16.717}}} (0.058) \\
        1000 & \textcolor{black}{32.200} (0.463)  & \textcolor{black}{\textbf{\tsunderbarcomptime{29.770}}} (0.059) \\
        \hline
    \end{tabular}
\end{tabular}
\label{tab:comptime}
\end{table*}
}
\tsshort{%
\newcommand{\tsunderbarcomptime}[1]{{#1}}

\begin{table}[t]
    \caption{Computational times on Indian dataset.}    
    \centering
    \footnotesize
    \begin{tabular}[t]{|c|c|c|}
        \hline
        $n$ & Asym Tobit & Sym Tobit\\
        \hline
        10 & \textcolor{black}{0.330} (0.001)  & \textcolor{black}{\textbf{\tsunderbarcomptime{0.321}}} (0.001) \\
        17 & \textcolor{black}{0.536} (0.001)  & \textcolor{black}{\textbf{\tsunderbarcomptime{0.529}}} (0.001) \\
        31 & \textcolor{black}{\textbf{\tsunderbarcomptime{0.957}}} (0.006)  & \textcolor{black}{\tsunderbarcomptime{0.964}} (0.012) \\
        56 & \textcolor{black}{\textbf{\tsunderbarcomptime{1.706}}} (0.002)  & \textcolor{black}{1.715} (0.003) \\
        100 & \textcolor{black}{\tsunderbarcomptime{3.026}} (0.010)  & \textcolor{black}{\textbf{\tsunderbarcomptime{3.025}}} (0.002) \\
        177 & \textcolor{black}{\textbf{\tsunderbarcomptime{5.315}}} (0.011)  & \textcolor{black}{\tsunderbarcomptime{5.322}} (0.001) \\
        316 & \textcolor{black}{\textbf{\tsunderbarcomptime{9.429}}} (0.034)  & \textcolor{black}{\tsunderbarcomptime{9.434}} (0.032) \\
        562 & \textcolor{black}{\tsunderbarcomptime{16.665}} (0.046)  & \textcolor{black}{\textbf{\tsunderbarcomptime{16.633}}} (0.021) \\
        1000 & \textcolor{black}{\textbf{\tsunderbarcomptime{29.610}}} (0.094)  & \textcolor{black}{\tsunderbarcomptime{29.660}} (0.080) \\
        \hline
    \end{tabular}
    \label{tab:comptime-indian}
\end{table}
}
\section{Simulations}
\label{s:exp}
We carried out simulations to investigate the performance of the three correlation analysis methods described in Section~\ref{s:pccest}. Three water quality datasets including an Indian water dataset, a water dataset on NY Harbor, and a water dataset on Sapporo were used. The Indian water dataset contained 1,580 data, each of which contained six variates, FC, TC, pH, Cond, N, and BOD. The NY Harbor water dataset contained 292 records, each consisting of four variates, FC, TC, WT, and pH. These two datasets are available from https://www.kaggle.com/. The Sapporo water dataset is also publicly available from the supplement of Kato et al.'s paper~\cite{KatKobOis19}. The Sapporo dataset had 175 data, each of which included eight variates, E.coli, TC, pH, EC, SS, TN, TP, and FR. All of these have no detection limit. To simulate censoring situations, we chose two variates to regard as concentrations of two microorganisms. A virtual detection limit was assumed for each of the two microorganisms. The detection limits were selected so that the negative ratio is 0.8 for both microorganisms. For each dataset, $n=50$ records were randomly selected, and the concentration data of the two microorganisms of interest were censored. Three correlation analysis approaches were applied to the data prepared in this way. The estimated PCC $\hat{R}\in\{R_{\text{na\"{i}ve}},R_{\text{sym}},R_{\text{asym}}\}$ was assessed by the absolute error from the PCC computed from uncensored data. Namely, the error was defined as $|\hat{R}-R(\y_{\text{a}},\y_{\text{b}})|$, where $\y_{\text{a}}\in\bR^{n}$ and $\y_{\text{b}}\in\bR^{n}$ are the concentration data before censoring for two microorganisms, respectively. This procedure was repeated 50 times to obtain 50 errors. 

\tslong{%
Table~\ref{tab:indian}, Table~\ref{tab:harbor}, and Table~\ref{tab:sapporo} %
}
\tsshort{%
Table~\ref{tab:indian} in this paper and
Table~\REFTABHARVORARXIV$\,$ and Table~\REFTABSAPPOROARXIV$\,$ in \cite{CaoKat-arxiv21a} %
}
report the average of the estimation errors for all choices of two microorganisms for the Indian, NY Harbor, Sapporo water datasets, respectively. The standard deviations are presented in parentheses. The minimal error among three errors in each row is bold-faced. 
Asym Tobit, Sym Tobit, and Na\"{i}ve denote the asymmetric Tobit, classical Tobit, and na\"{i}ve approaches, respectively. 
For the Indian dataset, the asymmetric and classical Tobit approaches obtained the minimal error for 28 pairs and two pairs, respectively, whereas the na\"{i}ve approach could not obtain the minimal error for any pair. For the NY Harbor dataset, the asymmetric and classical Tobit approaches achieved the minimal errors for 12 and two pairs, respectively. For Sapporo dataset, the two Tobit approaches obtained the minimal errors for 39 and 19 pairs. The na\"{i}ve approach did not obtain the minimal error for any pair. 
These results suggest that imputation of undetected observations leads to a better estimation performance for inferring PCC compared to the na\"{i}ve approach. Let us speculate why imputation leads to a better estimation. Common visible observations $\cI_{vv}$ tend to be few when two microorganisms are correlated poorly. In such a case, the na\"{i}ve method computes the PCC from a small, paired dataset, which worsens estimation. The empirical observations that the asymmetric Tobit performed better than the classical Tobit for many pairs suggested that the asymmetric Tobit exploited domain knowledge effectively to impute undetected concentrations, resulting in more precise estimations. 

Moreover, the runtimes for fitting the Tobit models are reported. A major technical contribution of this study is finding that the M-step of the EM algorithm is reduced to NNLS even when the prior of the regression coefficients is replaced with a slightly complicated distribution named the asymmetric normal distribution. This is in contrast to the EM algorithm for fitting the classical Tobit model, in which the M-step can be performed by solving an ordinary unconstrained least square problem. NNLS is a constrained convex program. Given this, how much additional runtime is required for fitting the asymmetric Tobit model, compared to fitting the symmetric Tobit model? To answer this question empirically, the runtimes of 30 iterations of the EM algorithms for the asymmetric and symmetric Tobit models were measured with a variable sample size $n$. 
\tslong{%
Table~\ref{tab:comptime}(a), (b), and (c) %
}
\tsshort{%
Table~\ref{tab:comptime-indian} in this paper and %
Table~\REFTABRUNTIMEARXIV$\,$(b),(c) in \cite{CaoKat-arxiv21a} %
}
report the average CPU times over 10 trials for the Indian, NY Harbor, and Sapporo water datasets, respectively. The unit is seconds in the tables. The figures in parentheses are the standard deviations. Surprisingly, no significant differences between the two models were observed even though the M-step for the asymmetric Tobit is a constrained least square problem because the number of regression coefficients is not very large in the application of water quality analysis. The dimensionalities for the three datasets, say $d$, were only six, four, and eight, respectively, thus solving  the NNLS problems quickly compared to the E-step in which the values of the cumulative density function are computed for $(n-n_{\text{v}})$ data. To examine the case where the number of regression coefficients is large, an artificial dataset was generated with $d=200$ and the computational times were examined. When $d$ was large, solving NNLS became a computationally expensive step, and thereby the differences between the computational times of the two models appeared clearly, as shown in 
\tslong{%
Table~\ref{tab:comptime}(d). %
}
\tsshort{%
Table~\REFTABRUNTIMEARXIV$\,$(d) in \cite{CaoKat-arxiv21a}. %
}
However, when the sample size $n$ was increased, the ratio of the two computational times approached one, because the computational cost of the E-step was again dominant with larger $n$. To summarize the results of our investigation of runtimes, we simply note the intended application of water quality analysis. In this application, the dimensionality $d$ is small, meaning that the additional computational cost paid for NNLS can be ignored.

\section{Conclusions}
In this paper, we demonstrated the favorable effects of imputation of undetected observations using side information prior to correlation computation for analysis of relationships between left-censored data pairs, with the aim of applying pathogenic concentration data to assess exposure risk to pathogens in water. The simulation results suggested that exploitation of domain knowledge for imputation of undetected data made the use of side information more effective. The asymmetric normal prior was introduced to the Tobit model as the key tool for imputation of undetected data. We showed theoretically that each iteration of the EM algorithm for Tobit fitting with the asymmetric prior can run efficiently by reducing the sub-problem for the M-step to the nonnegative least square problem, which is known as a quickly solvable convex problem. In future work, the method developed in this study will be applied to actual analysis of pathogen concentration data to redesign an improved routine to periodic monitoring, given that pathogen measurement technologies keep evolving. 
\section*{Acknowledgment}
    This research was performed by the Environment Research and Technology Development Fund JPMEERF20205006 of the Environmental Restoration and Conservation Agency of Japan and supported by JSPS KAKENHI Grant Number 19K04661.    

\renewcommand{\newblock}{}
\bibliographystyle{plain}

\begin{thebibliography}{10}

\bibitem{Amemiya1984}
Takeshi Amemiya.
\newblock Tobit models: A survey.
\newblock {\em Journal of Econometrics}, 24(1-2):3--61, January 1984.

\bibitem{Bellavia2006}
Stefania Bellavia, Maria Macconi, and Benedetta Morini.
\newblock An interior point newton-like method for non-negative least-squares
  problems with degenerate solution.
\newblock {\em Numerical Linear Algebra with Applications}, 13(10):825--846,
  2006.
\newblock doi: 10.1002/nla.502.

\bibitem{Boehm2014a}
A.~B. Boehm and L.~M. Sassoubre.
\newblock {\em Enterococci: From Commensals to Leading Causes of Drug Resistant
  Infection}, chapter Enterococci as Indicators of Environmental Fecal
  Contamination.
\newblock Massachusetts Eye and Ear Infirmary in Boston, editors: Michael S
  Gilmore, Don B Clewell, Yasuyoshi Ike, Nathan Shankar, 2014.

\bibitem{DonghuiChen09-nonneg}
Donghui Chen and Robert~J. Plemmons.
\newblock Nonnegativity constraints in numerical analysis.
\newblock In Adhemar Bultheel and Ronald Cools, editors, {\em The Birth of
  Numerical Analysis}, pages 109--139. World Scientific, Nov 2009.
\newblock doi: 10.1142/9789812836267\_0008.

\bibitem{Dobrowoski08a}
James Dobrowoski, Michael O'Neill, Lisa Duriancik, and Joanne Throwe.
\newblock Opportunities and challenges in agricultural water reuse: Final
  report.
\newblock {\em USDA-CSREES}, 89:--, - 2008.

\bibitem{GenVinLip09}
J.~Gentry, J.~Vinje, and E.~K. Lipp.
\newblock A rapid and efficient method for quantitation of genogroups i and ii
  norovirus from oysters and application in other complex environmental
  samples.
\newblock {\em J Virol Methods}, 156(1--2):59--65, Mar 2009.

\bibitem{GohSaeGu19}
S.~G. Goh, N.~Saeidi, X.~Gu, G.~G.~R. Vergara, L.~Liang, H.~Fang, M.~Kitajima,
  A.~Kushmaro, and K.~Y. Gin.
\newblock Occurrence of microbial indicators, pathogenic bacteria and viruses
  in tropical surface waters subject to contrasting land use.
\newblock {\em Water Res}, 150(-):200--215, Mar 2019.

\bibitem{ItoKatHasKatIshOkaSan17a}
Toshihiro Ito, Tsuyoshi Kato, Makoto Hasegawa, Hiroyuki Katayama, Satoshi
  Ishii, Satoshi Okabe, and Daisuke Sano.
\newblock Evaluation of virus reduction efficiency in wastewater treatment unit
  processes as a credit value in the multiple-barrier system for wastewater
  reclamation and reuse.
\newblock {\em Journal of Water and Health}, 14(5):879--889, Dec 2016.

\bibitem{ItoKatTak15a}
Toshihiro Ito, Tsuyoshi Kato, Kenta Takagishi, Satoshi Okabe, , and Daisuke
  Sano.
\newblock Bayesian modeling of virus removal efficiency in wastewater treatment
  processes.
\newblock {\em Water Science and Technology}, 72(10):1789--95, Nov 2015.
\newblock doi: 10.2166/wst.2015.402.

\bibitem{KatKobOis19}
T.~Kato, A.~Kobayashi, W.~Oishi, S.~S. Kadoya, S.~Okabe, N.~Ohta, M.~Amarasiri,
  and D.~Sano.
\newblock Sign-constrained linear regression for prediction of microbe
  concentration based on water quality datasets.
\newblock {\em J Water Health}, 17(3):404--415, Jun 2019.

\bibitem{KatKobIto15a}
Tsuyoshi Kato, Ayano Kobayashi, Toshihiro Ito, Takayuki Miura, Satoshi Ishii,
  Satoshi Okabe, and Daisuke Sano.
\newblock Estimation of concentration ratio of indicator to pathogen-related
  gene in environmental water based on left-censored data.
\newblock {\em Journal of Water and Health}, 14(1):14--25, Feb 2016.
\newblock doi:10.2166/wh.2015.029.

\bibitem{KatMiuOkaSan13a}
Tsuyoshi Kato, Takayuki Miura, Satoshi Okabe, and Daisuke Sano.
\newblock Bayesian modeling of enteric virus density in wastewater using
  left-censored data.
\newblock {\em Food and Environmental Virology}, 5(4):185--193, Dec 2013.

\bibitem{KatOmaAso02a}
Tsuyoshi Kato, Shinichiro Omachi, and Hirotomo Aso.
\newblock Asymmetric gaussian and its application to pattern recognition.
\newblock In {\em Joint IAPR International Workshops on Syntactical and
  Structural Pattern Recognition and Statistical Pattern
  Recognition(S+SSPR2002)}, pages 405--413, 2002.

\bibitem{KorMcMHar18}
A.~Korajkic, B.~R. McMinn, and V.~J. Harwood.
\newblock Relationships between microbial indicators and pathogens in
  recreational water settings.
\newblock {\em Int J Environ Res Public Health}, 15(12):--, Dec 2018.

\bibitem{KraHerCra-MCSS96a}
M.~H. Kramer, B.~L. Herwaldt, G.~F. Craun, R.~L. Calderon, and D.~D. Juranek.
\newblock Surveillance for waterborne-disease outbreaks--united states,
  1993-1994.
\newblock {\em MMWR CDC Surveill Summ}, 45(1):1--33, Apr 1996.

\bibitem{Lawson1995solving}
Charles~L. Lawson and Richard~J. Hanson.
\newblock {\em Solving Least Squares Problems}.
\newblock Society for Industrial and Applied Mathematics, jan 1995.
\newblock doi:10.1137/1.9781611971217.

\bibitem{McMinn-lam2017}
B.R. McMinn, N.J. Ashbolt, and A.~Korajkic.
\newblock Bacteriophages as indicators of faecal pollution and enteric virus
  removal.
\newblock {\em Letters in Applied Microbiology}, 65(1):11--26, June 2017.

\bibitem{Meinshausen2013}
Nicolai Meinshausen.
\newblock Sign-constrained least squares estimation for high-dimensional
  regression.
\newblock {\em Electronic Journal of Statistics}, 7:1607--1631, 2013.
\newblock doi: 10.1214/13-ejs818.

\bibitem{NapHonIch19}
S.~P. Nappier, T.~Hong, A.~Ichida, A.~Goldstone, and S.~E. Eftim.
\newblock Occurrence of coliphage in raw wastewater and in ambient water: A
  meta-analysis.
\newblock {\em Water Res}, 153(-):263--273, Apr 2019.

\bibitem{Noble2005}
Rachel~T. Noble and Stephen~B. Weisberg.
\newblock A review of technologies for rapid detection of bacteria in
  recreational waters.
\newblock {\em Journal of Water and Health}, 3(4):381--392, December 2005.

\bibitem{IWP04}
Committee on~Indicators~for Waterbone~Pathogens.
\newblock {\em Indicators for Waterborne Pathogens}.
\newblock National Academies Press, 2004.

\bibitem{Pedrero-AWM10a}
Francisco Pedrero, Ioannis Kalavrouziotis, Juan~Jose Alarcon, Prodromos
  Koukoulakis, and Takashi Asano.
\newblock Use of treated municipal wastewater in irrigated agriculture--review
  of some practices in spain and greece.
\newblock {\em Agricultural Water Management}, 97(9):1233--1241, Sept 2010.

\bibitem{SedVarMeo18}
M.~I. Sedji, M.~Varbanov, M.~Meo, M.~Colin, L.~Mathieu, and I.~Bertrand.
\newblock Quantification of human adenovirus and norovirus in river water in
  the north-east of france.
\newblock {\em Environ Sci Pollut Res Int}, 25(30):30497--30507, Oct 2018.

\bibitem{XiaotongWen-Sus20}
Xiaotong Wen, Feiyu Chen, Yixiang Lin, Hui Zhu, Fang Yuan, Duyi Kuang, Zhihui
  Jia, and Zhaokang Yuan.
\newblock Microbial indicators and their use for monitoring drinking water
  quality{\textemdash}a review.
\newblock {\em Sustainability}, 12(6):2249, March 2020.

\end{thebibliography}

\end{document}